\documentclass[%
 reprint,
superscriptaddress,
showpacs,preprintnumbers,
 amsmath,amssymb,
 aps,
]{revtex4-1}

\usepackage{graphicx}% Include figure files
\usepackage{dcolumn}% Align table columns on decimal point
\usepackage{bm}% bold math
\usepackage{color}

\begin{document}
\title{Quantum capacitance of double-layer graphene}

\author{Fariborz Parhizgar}
\affiliation{School of Physics, Institute for Research in
Fundamental Sciences (IPM), Tehran 19395-5531, Iran}

\author{Alireza Qaiumzadeh}
\affiliation{Department of Physics, Norwegian University of Science and Technology, NO-7491 Trondheim, Norway}

\author{Reza Asgari}
\email{asgari@ipm.ir}
\affiliation{School of Physics, Institute for Research in
Fundamental Sciences (IPM), Tehran 19395-5531, Iran}
\affiliation{School of Nano Science, Institute for Research in
Fundamental Sciences (IPM), Tehran 19395-5531, Iran}
\date{\today}
\begin{abstract}
We study the ground-state properties of a double-layer graphene system with the Coulomb interlayer electron-electron interaction modeled within the random phase approximation. We first obtain an expression of the quantum capacitance of a two-layer system. In addition, we calculate the many-body exchange-correlation energy and quantum capacitance of the hybrid double-layer graphene system at zero temperature. We show an enhancement of the majority density layer thermodynamic density of states owing to an increasing interlayer interaction between two layers near the Dirac point. The quantum capacitance near the neutrality point behaves like square root of the total density, $\alpha \sqrt{n}$, where the coefficient $\alpha$ decreases by increasing the charge density imbalance between two layers. Furthermore, we show that the quantum capacitance changes linearly by the gate voltage. Our results can be verified by current experiments.

\end{abstract}
\maketitle

\section{Introduction}

The conventional
double layers based on semiconductors have amassed
great interest for many years. Double layer electron (hole) systems comprise two parallel
quasi two-dimensional (2D) electron (hole) layers in close proximity~\cite{ref:Hanna}. Such systems are useful structures to study various novel
physical phenomena arising from the interlayer interaction effects specially at low particle densities or close proximity distance where many-body physics are significant. For instance, the observation of the fractional quantum Hall
state at a half filling factor which is forbidden in
monolayer structures~\cite{destruction_IQH, FQH}, the quantum Hall ferromagnetic phase transition
~\cite{ref:Yang}, the Coulomb drag in a double layer system~\cite{ref:Gramila}, the
quantum capacitance and the electronic compressibility and transport properties of bilayer structures are systems where the interlayer interactions give rise to new physical properties.

Graphene, a flat sheet of carbon atoms arranged in a honeycomb lattice~\cite{Novoselov,Castro Neto}, after realization in 2004, has been attracting the attention of many scientists in different research areas from both technological and academic points of view. The low-energy charge carriers in pristine graphene behave as massless Dirac fermions. Since the density of states of monolayer graphene changes linearly as the Fermi energy, therefore, the quantum capacitance, which is a consequence of the Pauli principle~\cite{Luryi} requires extra energy for filling a quantum system with electrons, can be changed by applying a gate voltage. The differential capacitance of graphene is linearly proportional to its electric potential when operated near the Dirac point. Recently, the local compressibility of graphene has been measured~\cite{martin} and is consistent with the many-body calculations~\cite{peres} of this quantity. Moreover, experiments ~\cite{yu} on measuring quantum capacitance in pristine graphene revealed the signature of many-body effects in agreement with theoretical calculations~\cite{asgari2014}. The quantum capacitance of graphene has been measured using a three-electrode electrochemical configuration~\cite{xia}, the graphene-insulator-semiconductor backgate~\cite{Giannazzo}, the metal-oxide-semiconductor structures~\cite{xu}, in double-layer capacitors~\cite{Ji} and the epitaxial graphene layers thermally elaborated on a carbon terminated face~\cite{Trabelsi}.  The quantum capacitance of a bilayer graphene has been measured~\cite{bilayer_exp} and studied theoretically~ \cite{bilayer_the} too. Furthermore, the ground-state properties and dynamical behavior of gapless and gapped graphene monolayers have been the subject of many theoretical studies~\cite{Castro Neto, Alireza, Pyatkovskiy}.

\begin{figure}
\includegraphics[width=0.9\linewidth]{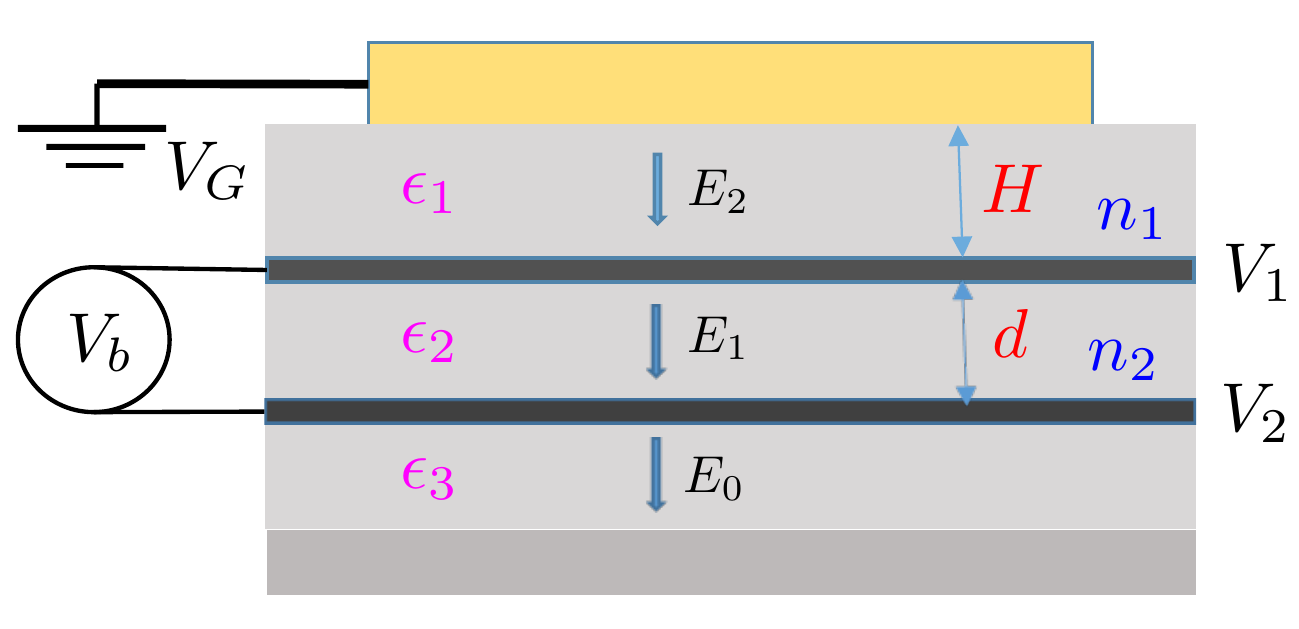}
\caption{(Color online) Schematic of a double layer system, including two graphene layers encapsulated by different materials and separated by a distance $d$ from each other. The spacer distance is $H$ and the gate voltage is applied after the spacer material. The bias voltage and the charge carrier densities together with corresponding voltages of each layer $V_b$, $V_1$ and $V_2$, respectively, are shown. Applying an electric field $E_2$ on the top layer results a penetrating field $E_1$ between layers. By connecting the bottom layer to a large resistance, the electric field, $E_0$ is fixed, and then the physical quantity $R_{E}=dE_1/dE_2$ can be obtained.
\label{Schem}}
\end{figure}

Multilayer systems enhance the effects of interparticle interactions through the combination of reduced dimensionality and low particle density and by the layer index. In this article, we study the quantum capacitance of a double-layer graphene with imbalanced charge densities by using the screened Coulomb potentials within the random phase approximation (RPA) to explore the effect of the interlayer interaction on the quantum capacitance.
The exchange-correlation energy are investigated previously for a double quantum well~\cite{Lian Zheng} and double-layer graphene~\cite{Profumo}.

One of the main consequences of the electron-electron interaction in graphene is the enhancement of the renormalized Fermi velocity, especially at lower densities~\cite{Borghi}. In contrast to massless Dirac fermions, the electron-electron interaction will result in the decreasing of the Fermi velocity in conventional 2D electron systems. We perform a theoretical study of the quantum capacitance and related quantities of a double layer system.  First of all, we generalize the expression of the quantum capacitance for a double layer structure and afterwards by using the ground-state properties of the system, we numerically calculate the quantum capacitance. The determination of graphene quantum capacitance is of crucial interest because not only does it get access to the density of states directly, but also an efficient way to explore various anomalies occurring near the Dirac point which might be difficult to probe only by transport measurement~\cite{Vignale,capacitance,Eisenstein,Jungwirth,ref:Hanna}. We show an enhancement of the majority density layer thermodynamic density-of-states owing to a reduction of the total electron density and thus an increasing interlayer interaction between two layers near the Dirac point. Our numerical results show that the quantum capacitance near the neutrality point behaves like a square root of the total density, $\alpha \sqrt{n}$, where the coefficient $\alpha$ decreases by increasing the charge density imbalance between two layers.  We also find that the quantum capacitance is sensitive to the bias voltage and depends on the gate voltage linearly, in contrast to a conventional 2D electron gas system where its quantum capacitance is
independent of gate voltage.

This paper is organized as follows. In Sec. II we present our model, discuss the realization of our set-up and study the effect of the Coulomb interactions on the ground-state energy of the double layer systems. We then discuss the quantum capacitance for the double-layer graphene systems. The numerical results and discussion are presented in Sec. III and finally we conclude our results in Sec. IV with a summary and some remarks.

\section{Theory and Model Hamiltonian}
We consider a free-disordered double-layer structure incorporating a doped graphene layer (layer I) placed on another graphene layer (layer II) with a separation distance $d$ at zero-temperature. A schematic of the structure is shown in Fig.~\ref{Schem}, where the electron densities in the layers are $n_1$ and $n_2$. We assume that each layer is about zero thickness in the direction normal to the plane of the system. The layers are separated by a dielectric material (shown in Fig.~\ref{Schem}) with a dielectric constant $\epsilon_2$ and we suppose that the tunneling of electrons between the layers is negligible, however, the Coulomb interlayer interaction plays an important role in the system. The effective Hamiltonian of the system under consideration reads

\begin{align}\label{eq:Hamiltonian}
\hat{\mathcal{H}}=&\sum_{{\bf{k}},l,\alpha,\beta}\hat{\psi}^\dag_{{\bf{k}},l,\alpha}
\hbar
v_l{\bm{\sigma}}_{\alpha,\beta}\cdot{\bf{k}}\hat{\psi}_{{\bf{k}},l,\beta}\nonumber\\&+
\frac{1}{2\mathcal{A}}\sum_{{\bf{q}}\neq 0,l,l'}
{\bf V}_{ll'}(q)\hat{\rho}_{{\bf{q}},l}\hat{\rho}^\dag_{{\bf{q}},l'}.
\end{align}

Here $v_l$ is the Fermi velocity of layer $l=1$ and $2$, $\hat{\psi}_{{\bf{k}}}$'s are  the corresponding two-component pseudospinors of the noninteracting Hamiltonian, $\mathcal{A}$ is the area of the system, ${\bf V}_{ll'}$ is the matrix of the bare
Coulomb interactions, $\hat{\rho}_{{\bf{q}},l}$ is the density
operator for the $l-$th layer, $\alpha$ and $\beta$ are the pseudospin
indeces, and $\bm{\sigma}$ is the Pauli matrices. In this Hamiltonian, the two
layers are perfectly decoupled, and the long-range Coulomb
interaction affects only the electrons in the layers.
We define the retarded
density-density linear response functions
$\chi_{ll'}({\bf{q}},\omega)=\frac{1}{i\hbar}\lim_{\eta\rightarrow
0}\int_0^{\infty} e^{i(\omega+i\eta)
t}\langle[\hat{\rho}_l({\bf{q}},t),\hat{\rho}^\dag_{l'}({\bf{q}})]\rangle$,
where $\eta$ is an infinitesimal parameter, and $\langle ...
\rangle$ denotes the average in the thermal equilibrium ensemble. Notice that in
our model, $\chi_{ll'}=\chi_l\delta_{l,l'}$. This quantity is related
to several important many-body properties, such as the total ground-
state energy, the electron compressibility, and the renormalized
Fermi velocity~\cite{Barlas}. Using the RPA, one can
find ${\bf \chi}^{-1}({\bf{q}},\omega)={\bf \chi}^{(0)^{-1}}({\bf{q}},\omega)-{\bf V}(q)$,
where ${\bf \chi}^{(0)}({\bf{q}},\omega)$ is the noninteracting Lindhard matrix
response function of the Dirac fermions in the two-component systems. $V(q)$ is a $2\times2$ matrix including inter- and intralayer Coulomb interactions.

The intra- and interlayer Coulomb potentials are given by
\begin{eqnarray}
V_{11}(q)&=&\frac{4\pi e^2}{q F(q)}[(\epsilon_2
+\epsilon_3)e^{qd}+(\epsilon_2 -\epsilon_3)e^{-qd}],\nonumber\\
V_{12}(q)&=&V_{21}(q) =\frac{8\pi e^2}{q F(q)} \epsilon_2,
\end{eqnarray}
where $F(q)=(\epsilon_1 +\epsilon_2)(\epsilon_2
+\epsilon_3)e^{qd}+(\epsilon_1 -\epsilon_2)(\epsilon_2
-\epsilon_3)e^{-qd}$, and the interaction in the top layer can be
obtained by replacing $\epsilon_1 \leftrightarrow \epsilon_3$ in
$V_{11}(q)$. Here $\epsilon_i$ is the dielectric constant of the region
$i$th.

The exchange and the correlation energies of a double layer system in the RPA approximation
have been calculated in Ref.~[\onlinecite{Profumo}]. We follow that approach to calculate the ground-state energy of the system under consideration.
We define the total Fermi wave vector of
the system as $k_{\rm F}=\sqrt{4\pi n/g}$ ($n=n_1+n_2$) where $g(=g_sg_v)$ is the total spin ($g_s$), and valley ($g_v$)
degeneracies.

Having calculated the ground-state energy, the electron compressibility and the quantum capacitance ~\cite{Luryi} can be obtained which are two important physical quantities to investigate many-body effects in 2D systems. These quantities are in fact related to the density-of-states.

The electron compressibility of a system~\cite{Borghik} is related to the total
energy as
\begin{equation}\label{eq:compressibility}
\kappa ^{-1}=n^2\frac{\partial \mu}{\partial n}=n^2\frac{\partial^2
(n {\varepsilon})}{\partial n^2},
\end{equation}
where $\mu$ is the chemical potential, and the total energy per particle is
${\varepsilon}=\varepsilon_{kin}+\varepsilon_{x}+\varepsilon_{c}$,
where $\varepsilon_{kin}$ is the total kinetic energy of the system: $\varepsilon_{kin} = (g \varepsilon_{\rm F} /6)[\bar{k}_{\rm F2}^3 + \bar{k}_{\rm F1}^3 ]$ where $k_{{\rm F}i}=\bar{k}_{{\rm F}i} k_{\rm F}$ denotes the Fermi wavelength of the layer $i$ and the energy quantities are scaled in units of the $\varepsilon_{\rm F}=\hbar v_{\rm F}k_{\rm F}$. We assume that the density-of-states is symmetric with respect to the Fermi energy.

Eisenstein {\it et al}.~\cite{Eisenstein} have introduced a powerful method to measure the compressibility of a 2DEG layer by locating another 2D layer in the proximity of the first layer. The spacer distance is $H$ and the gate voltage is applied after the spacer material (see Fi. 1). The bias voltage and the charge carrier densities together with corresponding voltages of each layer $V_b$, $V_1$ and $V_2$, are considered, respectively. Applying an electric field $E_2$ on the top layer results a penetrating field $E_1$ between layers. By connecting the bottom layer to a large resistance, the electric field, $E_0$ is fixed, and then the physical quantity $R_{E}=dE_1/dE_2$ can be measured.
Before their experimental proposal, the measurement of the compressibility was based on measuring the capacitance between a 2DES layer and a metal gate. In this method, the capacitance was obtained by the sum of a large geometric contribution and a much smaller term $\partial \mu /\partial n$. It turns out that the first term made significant errors in the measurement. In the method of Eisenstein {\it et al}. method, a double quantum well system was used as shown in the Fig.~\ref{Schem}. Using a large resistance connected to the bottom layer, one can fix the value of $E_0$. In addition, tuning the external field $E_2$ results in the field changes between layers $E_1$ and therefore, $dE_1/dE_2$ is the quantity which can be measured. Eisenstein {\it et al}. used this method in a situation where the distance between layers was far enough so the interlayer correlations can be neglected, and then measured the compressibility of the top layer. Jungwirth and MacDonald~\cite{Jungwirth}, on the other hand, extended their analysis for a narrow double layer system where interlayer correlations play an important role.
\begin{figure}
\includegraphics[width=1\linewidth]{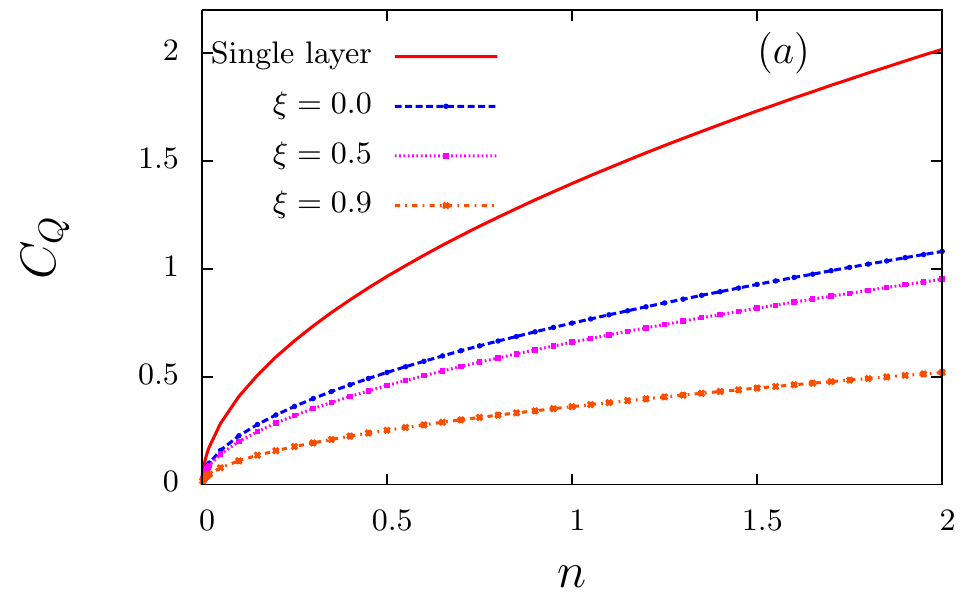}
\includegraphics[width=1\linewidth]{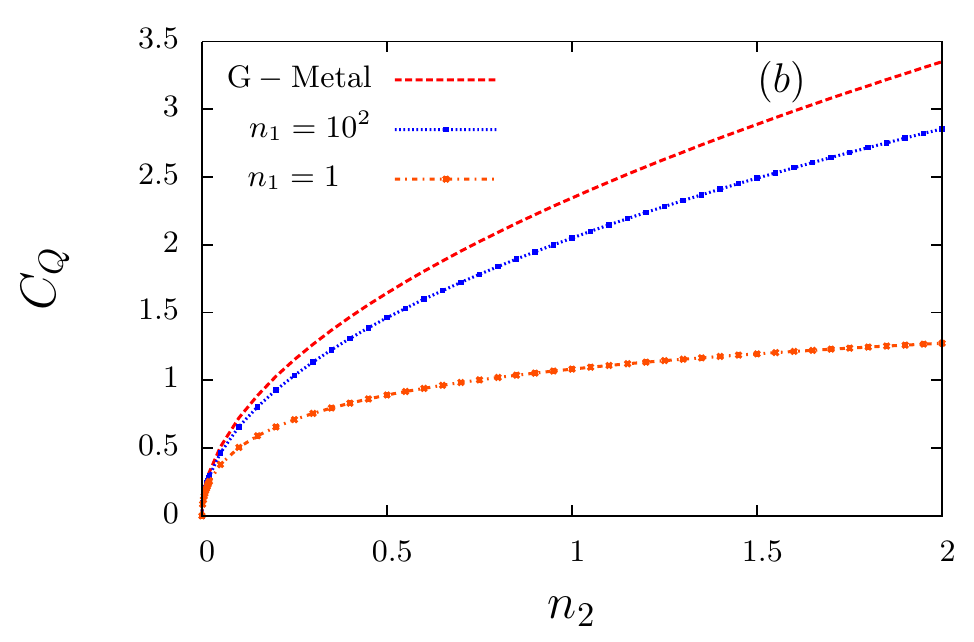}
\caption{(Color online) (a): The quantum capacitance (in units of $ {\rm
\mu F} {\rm cm}^{-2}$) from Eq.~(\ref{QC}) as a function of the electron density $n=n_1+n_2$
(in units of $10^{12}~{\rm cm}^{-2}$) for different value of the density imbalance, $\xi=(n_2-n_1)/n$. Here we consider $d=1$nm, $\epsilon_1=1$ with $H$ tends to infinity, $\epsilon_2=4.5$ and $\epsilon_3=4.5$. The solid curve shows the quantum capacitance of a single layer graphene. (b): The quantum capacitance (in units of $ {\rm
\mu F} {\rm cm}^{-2}$) as a function of the electron density $n_2$
(in units of $10^{12}~{\rm cm}^{-2}$) for different $n_1$ values (in units of $10^{12}~{\rm cm}^{-2}$) in comparison with that in a system incorporates encapsulated graphene with hBN and placed by a metal at a distance $d=1$nm studied in Ref.~[\onlinecite{asgari2014}].
\label{fig2}}
\end{figure}

\begin{figure}
\includegraphics[width=1\linewidth]{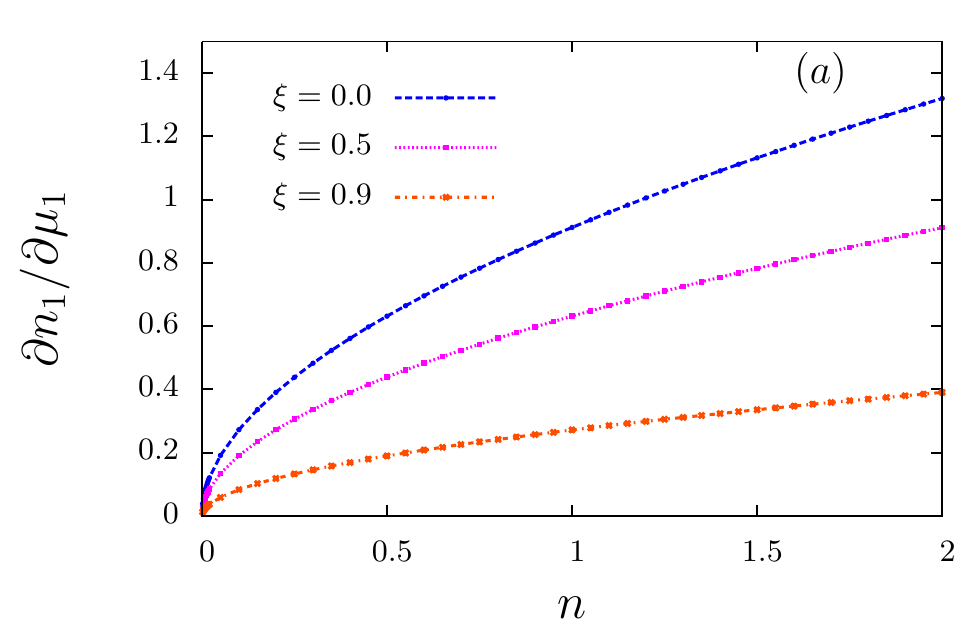}
\includegraphics[width=1\linewidth]{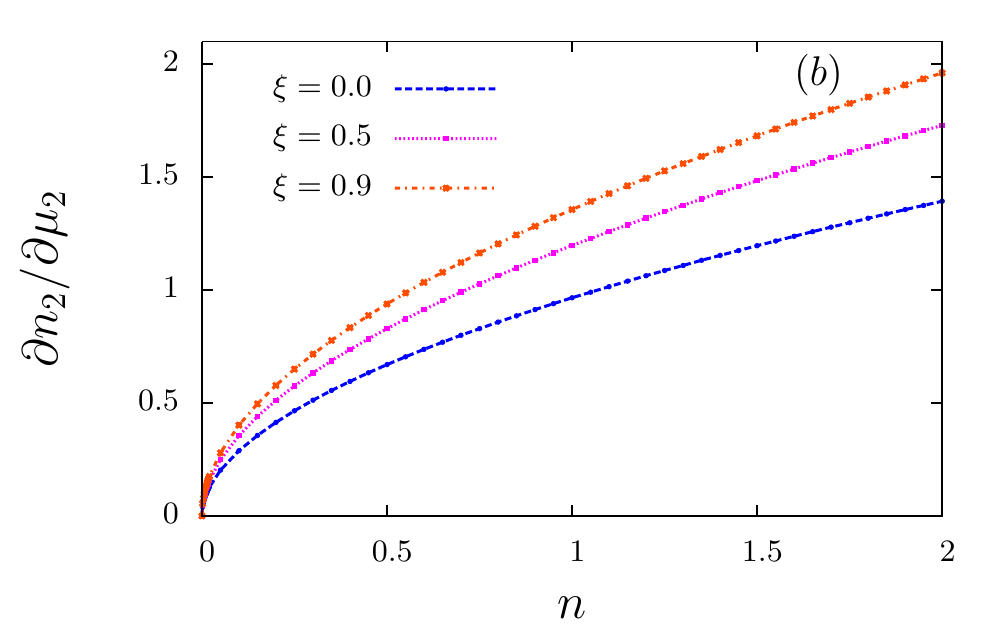}
\caption{(Color online) The thermodynamic density-of-states $\partial n_i/\partial \mu_i$ (in units of $10^{10}/({\rm meV} {\rm cm}^{2})$) where $i=1$ or $2$ as a function of the electron density $n=n_1+n_2$
(in units of $10^{12}~{\rm cm}^{-2}$) for different value of the density imbalance. We consider $d=1$nm, $\epsilon_1=1$ with $H$ tends to infinity, $\epsilon_2=4.5$ and $\epsilon_3=4.5$.
\label{fig3}}
\end{figure}

\begin{figure}
\includegraphics[width=1\linewidth]{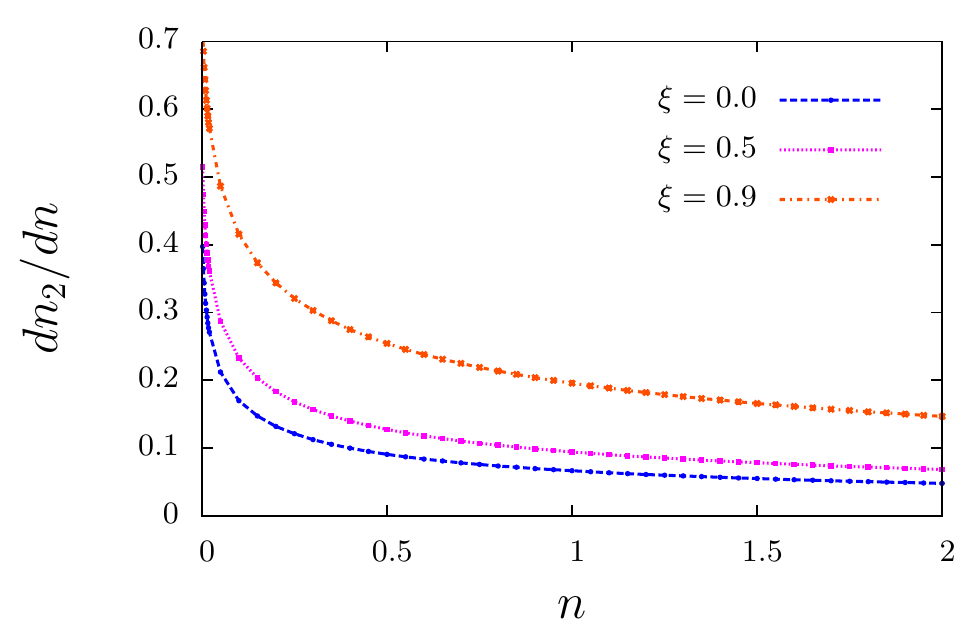}
\caption{(Color online) The Eisenstein ratio $d n_2/d n$ from Eq.~(\ref{dn}) as a function of the electron density $n=n_1+n_2$
(in units of $10^{12}~{\rm cm}^{-2}$) for different value of the density imbalance and $d=1$nm, $\epsilon_1=1$ with $H$ tends to infinity, $\epsilon_2=4.5$ and $\epsilon_3=4.5$.
\label{fig4}}
\end{figure}
The energy of the double layer system in this configuration can be
written up to an irrelevant constant by
\begin{eqnarray}\label{eq:capacity energy}
\frac{E(n_1,n_2)}{\mathcal{A}}=\frac{e^2 d}{2\epsilon_2} (n_2-n_0)^2+n\varepsilon (n-n_2,n_2)
\end{eqnarray}
where the first term is the energy stored in a capacitor with two parallel metallic planes, $n_2$ and $n_1$ are the densities of the bottom and top layers,
respectively, $n_0$ is a constant density, which is determined by $E_0$,
and $\varepsilon(n_1,n_2)$ is the energy per particle of the double-layer system.
From Poisson's equations and the electrostatic laws, we have
\begin{eqnarray}
\epsilon_2 E_1&=&\epsilon_3 E_0 -en_2, \nonumber\\
\epsilon_1 E_2&=&\epsilon_3 E_0 -e(n_1+n_2)
\end{eqnarray}
Changing $E_2$ varies the total density, whereas $E_1$ controls the electron density
$n_2$. Keeping in mind that $E_0$ is a constant value according to the large resistance and independent of the voltage, we thus define the Eisenstein ratio as $R_E=dE_1/dE_2=(\epsilon_2 / \epsilon_1)dn_2/dn$.
Having given $n$, we can determine $n_2$ by minimizing the total energy, using Eq. (\ref{eq:capacity
energy}) with respect to $n_2$. We thus get the following equation $\mu_1=\mu_2+(e^2d/\epsilon_2)(n_2-n_0)$ where $\mu_i=\partial [n \varepsilon(n_1,n_2)]/\partial n_i$. Note that $\mu_i$ includes all contributions to the chemical potential for electrons in the $i$th layer except for contributions from the electrostatic potentials and would be the full chemical potential if neutralizing positive charges in each layer were assumed.

It follows from $\mu_i $ by taking the partial derivatives with respect to the density of the layers that
\begin{eqnarray}
&&\frac{\partial \mu_1}{\partial n_1} dn_1+ \frac{\partial \mu_1}{\partial n_2} dn_2=\frac{\partial \mu_2}{\partial n_1} dn_1+ \frac{\partial \mu_2}{\partial n_2} dn_2+\frac{e^2d}{\epsilon_2}dn_2
\end{eqnarray}
and it turns out that
\begin{eqnarray}
(d_{11}-d_{21})(dn-dn_2)=(d+d_{22}-d_{12})dn_2
\end{eqnarray}
where we introduce a set of lengths as follows
\begin{eqnarray}
d_{ij}=\frac{\epsilon_2}{e^2}\frac{\partial \mu_i}{\partial n_j}.
\end{eqnarray}
where $i,j=1,2$ are layer labels. Accordingly, we find
\begin{eqnarray}\label{dn}
\frac{dn_2}{dn}=\frac{d_{11}-d_{21}}{d+d_{11}+d_{22}-d_{21}-d_{12}}.
\end{eqnarray}

Note that for the local minimum of the total energy per unit area, we require that the second derivative of the total energy, Eq. (\ref{eq:capacity energy}) with respect to $n_2$ be positive. This gives a necessary condition for stability in which $d+d_{11}+d_{22}-d_{21}-d_{12}>0$.

Moreover, the Eisenstein ratio~\cite{Hanna} is defined by $R_E=\epsilon_2 C/\epsilon_1 C_1$ where $C_1=\mathcal{A} e^2 \partial n_1/\partial \mu_1$ and thus
\begin{eqnarray}
%\frac{d_{11}-d_{21}}{d+d_{11}+d_{22}-d_{21}-d_{12}}=\frac{C d_{11}}{\epsilon_2 \mathcal{A} },\nonumber\\
\frac{1}{C}=\frac{d_{11}}{\epsilon_2\mathcal{A}}\frac{d+d_{11}+d_{22}-d_{21}-d_{12}}{d_{11}-d_{21}}.
\end{eqnarray}

It is easy to decompose the above expression into two parts as
\begin{equation}
\frac{1}{C}=\frac{1}{\epsilon_2\mathcal{A}}\frac{d}{1-\frac{d_{21}}{d_{11}}}+\frac{1}{\epsilon_1\mathcal{A}}\frac{d_{11}+d_{22}-d_{21}-d_{12}}{1-\frac{d_{21}}{d_{11}}}.
\end{equation}
Since $d_{21}\ll d_{11}$, we can approximate the denominator by unity and thus the capacitance $C$, which contains two contributions in the series, is
\begin{eqnarray}
\frac{1}{C}=\frac{1}{C_g}+\frac{1}{C_Q},
\end{eqnarray}
where $C_g=\mathcal{A}\epsilon_2/d$ is the geometrical capacitance and the quantum capacitance can be defined as
\begin{eqnarray}\label{QC}
C_Q=\mathcal{A}\frac{e^2}{\partial \mu_1/\partial n_1+\partial \mu_2/\partial n_2-\partial \mu_1/\partial n_2-\partial \mu_2/\partial n_1}
\end{eqnarray}
This expression of the quantum capacitance is a main equation in this paper. In the classical limit where $\hbar \rightarrow 0$ and $m \rightarrow \infty$ one finds $C_Q \rightarrow \infty$ and finally $C=C_g$. However, in the quantum regime, one can expect
interesting effects when $C_Q$ becomes compatible to the geometrical capacitance. Furthermore, by considering $d \rightarrow \infty$, this expression reduces to the well-known expression of the monolayer quantum capacitance where $C_Q=\mathcal {A }e^2\frac{\partial n_2}{\partial \mu_2}$.

\section{Numerical results}

In this section, we present our numerical results of the quantum capacitance using the ground-state energy at zero temperature through Eqs. 8-13. Here we introduce a density asymmetry parameter between two layers, $-1<\xi=(n_2-n_1)/n<1$.

Exchange and correlation energies in the system depend both on interactions on the Fermi wavelength scale which influence correlations between carriers and on interactions at shorter length scales which influence correlation with the Dirac sea background~\cite{Profumo}. Because decoupled graphene layers are separated by atomic length scales and carrier densities are always small, thus the atomic scale $k_{{\rm F},l}d$ is typically small. Here, the exchange energies are positive~\cite{Profumo} because they are calculated relative to zero carrier density using the Dirac point self-energy of this limit as the zero of energy and owing to its chirality. The increase in the exchange energy with the carrier density in graphene has the physical consequence of an enhancement of the screening and therefore, an increasing the renormalized Fermi velocity, instead of weakening it as in an ordinary 2D electron gas. The correlation energy, which is negative, is dramatically higher in the decoupled graphene layer in comparison to that in monolayer graphene, strongly influenced by interlayer interactions and resulting in the increased quantum capacitance.

Having calculated the ground-state energies based on the method reported in Ref. [\cite{Profumo}], we can calculate some interesting transport properties. The quantum capacitance as a function of the total electron density on the two layers is presented in Fig.~{\ref{fig2}}(a). The quantum capacitance starts from zero, in clean systems, at the Dirac point and increases by increasing the electron density owing to the fact that the renormalized Fermi velocity decreases. The behavior of $C_Q$ near the neutrality point is no longer linear and can be fitted quite well with the total density as $\alpha \sqrt{n}$ where the coefficient $\alpha$ decreases by increasing the density imbalance, $\xi$. These are happening owing to the change in the Fermi velocity stemming from the increasing of the Coulomb interlayer interaction~\cite{Profumo}. Our numerical results show that the quantum capacitance is suppressed by increasing the charge imbalance between two layers. We also calculate a quantum capacitance of a monolayer layer system, (the same as in Fig.~\ref{Schem} when $d\rightarrow \infty$) and the result is shown in Fig.~{\ref{fig2}}(a) as a solid line. In order to perceive the role of the Coulomb interactions, we compare the results with that in a system when a metal gate exists on the top of the encapsulated monolayer graphene [\onlinecite{asgari2014}]. The quantum capacitance as a function of the electron density $n_2$ for different $n_1$ is illustrated in Fig.~\ref{fig2}(b). First of all, there is an enhancement of the quantum capacitance when a metal is located close to the system. Furthermore, by reducing $n_1$, $\partial \mu_1/\partial n_1$ increases and thus $C_Q$ decreases. Therefore, by increasing $n_1$, the quantum capacitance of the double-layer graphene approaches the quantum capacitance of a graphene system in the presence of the metal gate.

The thermodynamic density-of-states $\partial n_i/\partial \mu_i$ for layer $i=1, 2$ as a function of the total electron density $n$ are shown in Fig.~\ref{fig3}. The thermodynamic density of states decreases with increasing $\xi$ for layer $I$ with the minority electron density $n_1=n(1-\xi)/2$, the $\partial n_2/\partial \mu_2$ increases, therefore, the interlayer interaction contribution of the minority density is dominated and results in increasing the thermodynamic density of states in the second layer. In the absence of interlayer interactions, $d_{12}=d_{21}=0$ and thus the $d_{ii}$ is related directly to the electronic compressibility in layer $i$.

\begin{figure}
\includegraphics[width=1\linewidth]{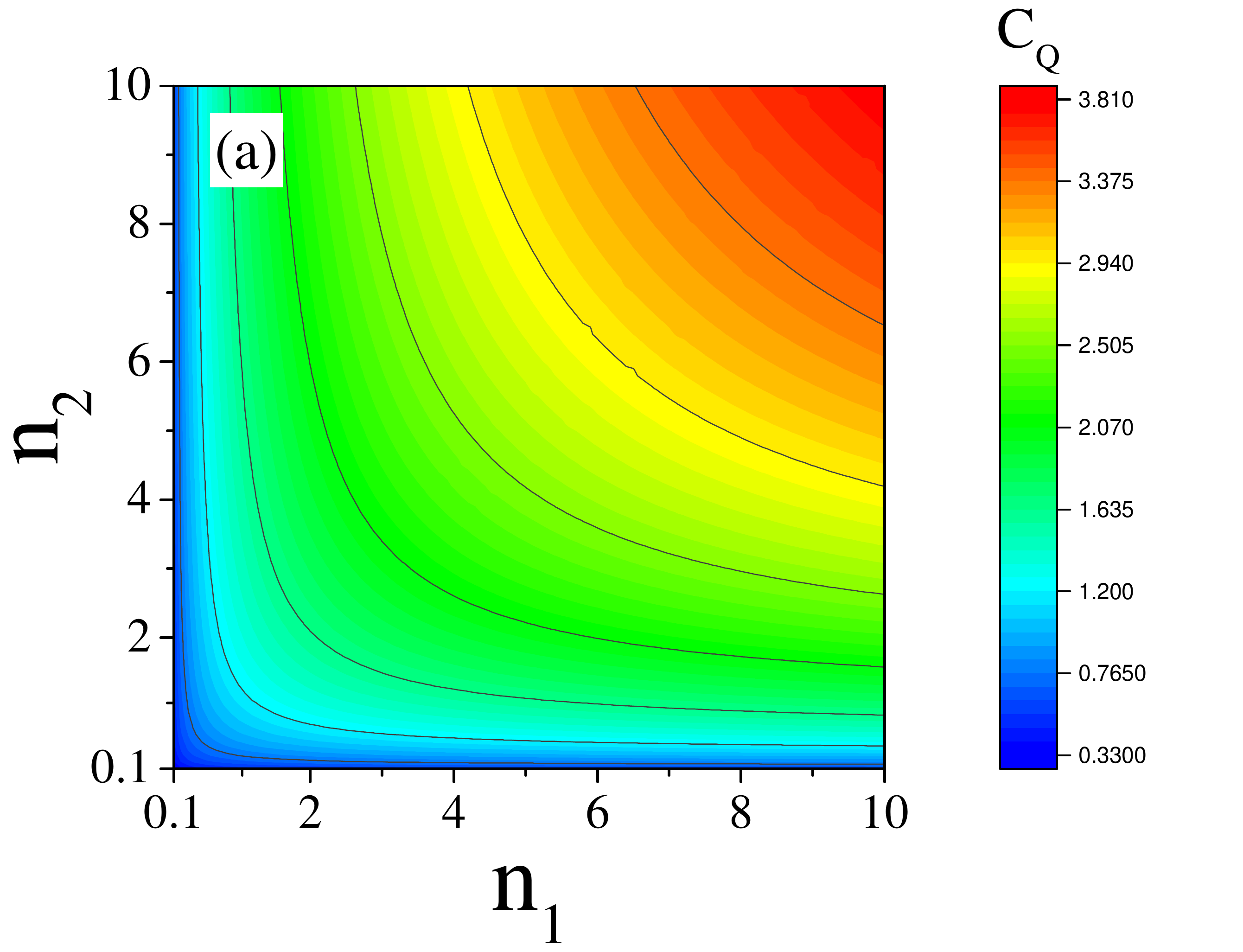}
\includegraphics[width=1\linewidth]{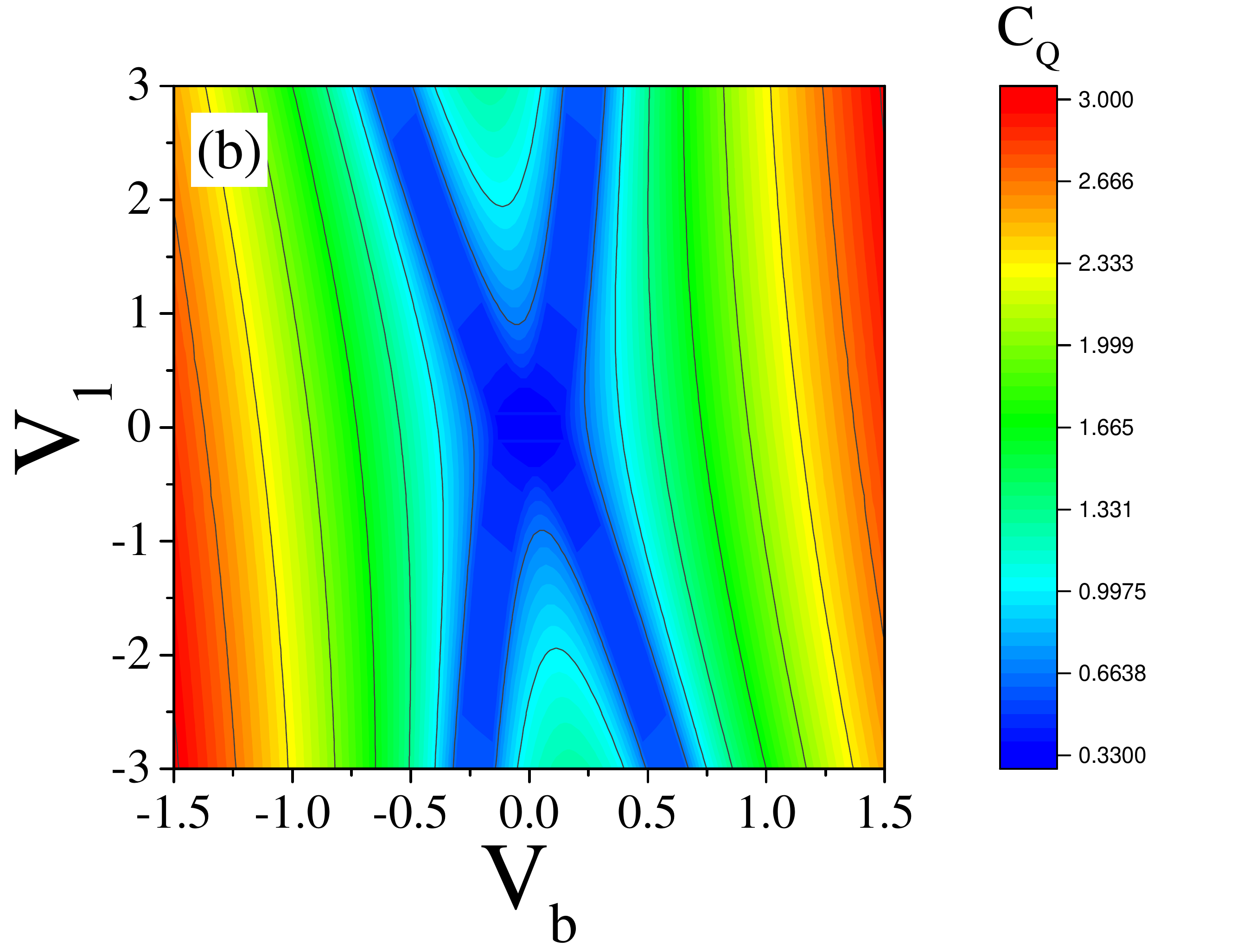}
\caption{(Color online) The quantum capacitance (in units of $ {\rm
\mu F} {\rm cm}^{-2}$) as functions of (a) the electron densities $n_1$ and $n_2$
(in units of $10^{12}~{\rm cm}^{-2}$) and (b) the top gate and bias potential in units of eV for $d=3.5$ and $H=30$ nm when graphene layers are encapsulated by hexagonal boron nitride where $\epsilon_i=4.5$ for $i=1$, 2 and 3. Notice that since $H/d$ is very large, the formalism given by Eqs. 1 and 2 are valid.
\label{fig5}}
\end{figure}

In Fig.~\ref{fig4}, we demonstrate the Eisenstein ratio, $dn_2/dn$ as a function of the total density $n$. The advantage of measuring $dn_2/dn$ is that this quantity depends only on the electronic lengths, $d_{ij}$ and the interlayer distance $d$. By increasing the electron density imbalance, the $d_{11}$ increases as well as $R_E$.

Finally, in order to connect this study to an experimental setup, we change the charge carrier densities or the top and bottom chemical potentials, $\mu_1$ and $\mu_2$ respectively, to the gate voltage $V_G$ and the interlayer bias $V_b$ assuming that all quantities are uniform in the horizontal directions~\cite{Profumo, Karina}. The bias voltage can be defined by the difference between the electrochemical potential of the top and bottom layers and reads as
\begin{equation}\label{vb}
V_b=V_1-V_2-(\mu_1-\mu_2)/e.
\end{equation}
Another electrostatic relation follows from the charge neutrality condition where $n_1+n_2+n_G=0$ where $n_G$ is the electron density associated with the gate voltage. These carrier densities are related to the electric fields (in what follows, we set $n_0$=0 and it means that all electric fields are measured with $E_0$)
\begin{eqnarray}
&E_2=-en_G/\epsilon_1,\nonumber\\
&\epsilon_2 E_1=\epsilon_1 E_2 -en_1. %\nonumber\\
%&\epsilon_1 E_2=\epsilon_3 E_0 -e(n_1+n_2),
\end{eqnarray}
Furthermore, the electric fields are connected to the voltages on the graphene layers and on the gate by
\begin{eqnarray}
&E_2=-(V_1-V_G)/H\nonumber\\
&E_1=-(V_2-V_1)/d
\end{eqnarray}
and the chemical potential for each layer can be calculated through $\mu_i=\partial[n\varepsilon(n_1,n_2)]/\partial n_i$ where $\varepsilon(n_1,n_2)$ is the ground-state energy per particle of the system calculated through the RPA. These aforementioned sets of equations can be solved for unknown $\mu_1$, $\mu_2$, $E_1$, $E_2$, $V_1$, $V_2$ and $V_G$ for given $n_1$ and $n_2$ values.

In Fig. 5(a) we plot the quantum capacitance in units of $ {\rm \mu F} {\rm cm} ^ {-2} $ as functions of the charge carrier densities $n_1$ and $n_2$, in units of $10^{12}~{\rm cm}^{-2}$, for $d=3.5$ and $H=30$ nm when graphene layers are encapsulated by hexagonal boron nitride. The $C_Q$ increases noticeably with increasing layers densities. Furthermore, the $C_Q$ is demonstrated as functions of the top gate $V_1$ and bias potential $V_b$ which is given by Eq.~\ref{vb} in Fig. 5(b). The domain of the potentials is considered in small regions. We find that the quantum capacitance is sensitive to the both potentials and it depends on the gate voltage linearly, on contrast to a conventional 2D electron gas system where its quantum capacitance is
independent of gate voltage. We stress that the implications of the findings reported in this work are applicable to any graphene devices that involve a metal-graphene contact. For example, for a graphene-based field-effect transistor, it has widely been reported that the contact resistance between the metal and the graphene exhibits an asymmetry with respect to the polarity of the gate potential~\cite{application,martin}.

\section{Summary and conclusion}
In conclusion, we have presented a theoretical scheme, based on the random phase approximation, to investigate the effects of Coulomb
interactions on observable quantities of decoupled graphene Fermi-liquid systems encapsulated by dielectric materials such as the quantum capacitance and charge compressibility
. We have carried out microscopic calculations to explore the ground-state properties of the system with the Coulomb interlayer electron-electron interaction modeled within the random phase approximation. Importantly, an expression for describing the quantum capacitance for a two layer system is derived. We have shown that the quantum capacitance near the neutrality point can be well behaved by $\alpha \sqrt{n}$ (linear behavior to the gate voltage) where the prefactor $\alpha$ decreases by increasing the charge density imbalance between two layers. An enhancement of the majority density layer thermodynamic density-of-states due to a reduction of the total electron density is calculated. We have also shown that the quantum capacitance increase by locating a metal on top of the considered system. Furthermore, we have found that the quantum capacitance is sensitive to the top and bottom potentials and the gate voltage dependence on the quantum capacitance is linear, in contrast to a conventional 2D electron gas system where its quantum capacitance is
independent of gate voltage.

\section{ACKNOWLEDGMENTS}

We would like to thank K. Novoselov and M. Polini for fruitful discussions. This work was partially supported by Iran Science Elites Federation grant no 11/66332.

\end{document}